%% file: ms.tex
\documentclass[journal=apchd5,manuscript=article]{achemso}
\usepackage{amssymb,epsfig,setspace}
\usepackage[version=3]{mhchem}
\pdfoutput=1
\graphicspath{{figs/}}

\input new

\title{Metal-Dielectric-Enhanced Upconversion: \\ Going ``Meso''}

\author{Ilia L. Rasskazov}
\affiliation{The Institute of Optics, University of Rochester, Rochester, New York 14627, United States}
\email{irasskaz@ur.rochester.edu}

\author{Alexander Moroz}
\affiliation{Wave-scattering.com}
\email{wavescattering@yahoo.com}

\author{Catherine J. Murphy}
\affiliation{Department of Chemistry, University of Illinois at Urbana-Champaign, Urbana, Illinois 61801, United States}

\author{Todd D. Krauss}
\affiliation{Department of Chemistry, University of Rochester, Rochester, New York 14627, United States}
\alsoaffiliation{The Institute of Optics, University of Rochester, Rochester, New York 14627, United States}
\alsoaffiliation{Materials Science Program, University of Rochester, Rochester, New York 14627, United States}

\author{P. Scott Carney}
\affiliation{The Institute of Optics, University of Rochester, Rochester, New York 14627, United States}

\begin{document}

\begin{tocentry}
\includegraphics[width=3.2in]{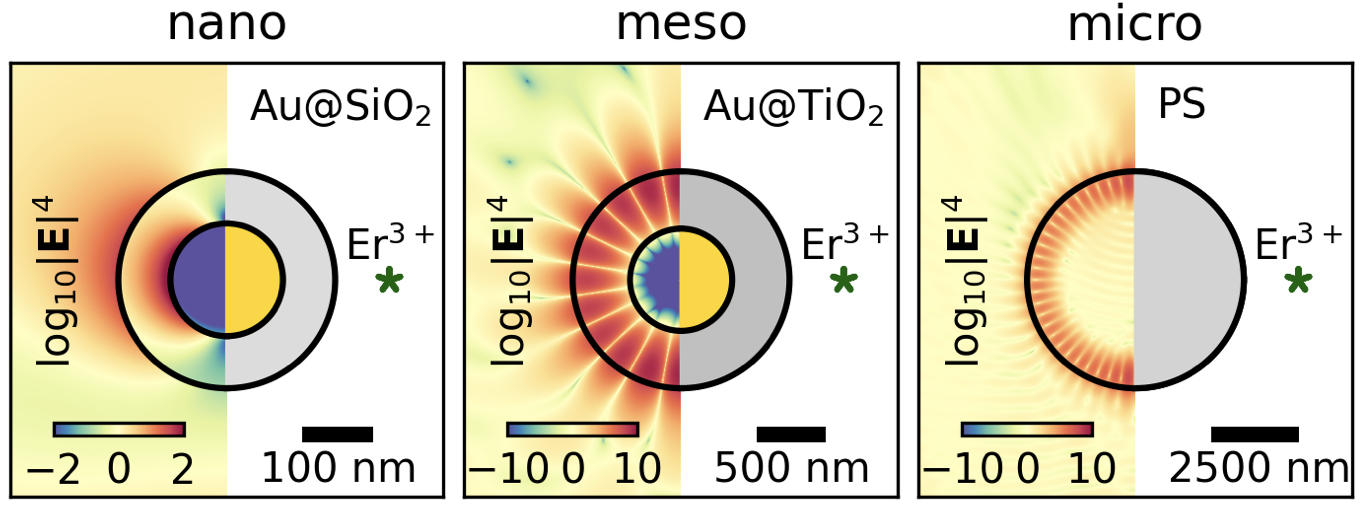}
\end{tocentry}

\begin{abstract}
A novel concept of a \textit{mesocavity}, based on a simple yet counter-intuitive idea of covering metal nanoparticles with unusually thick ($\gtrsim 100$ nm) high-index dielectric shells, is demonstrated in technologically relevant and largely sought-after case of an upconversion (UC) enhancement.
For readily available combinations of materials and geometrical parameters of metal-dielectric core-shell mesocavity, record high values of UC enhancement of $\gtrsim 10^6$ can be achieved on the cavity surface. 
For emitter inside the high-index dielectric shell with refractive index $n=2.7$, the Lorentz local-field correction can provide an additional enhancement factor for UC of $\approx 92$, resulting in UC enhancement of $\gtrsim 10^8$, which is three orders of magnitude larger than it has ever been reported for individual particles. 
Our mesocavity concept facilitates a beneficial synergy of plasmonic and whispering gallery-mode resonance functionalities within an intermediary region between nanoscale ($\lesssim 100$ nm) and microscale ($\gtrsim 1$ $\mu$m).
\end{abstract}

\section{Introduction}
Photon upconversion (UC), a sequential absorption of two or more low frequency photons and subsequent emission of light at higher frequency, has attracted significant interest in various fields, such as  biology~\cite{Chen2014,Chen2015,Zhou2015b,Wu2016,Wang2018d}, imaging~\cite{Huot2016}, solar energy harvesting~\cite{Wang2014,Jang2016,Meng2016,Li2018}, and nanoscopy~\cite{Liu2017f,Zhan2017}. 
Nonetheless, any wide-spread practical use of UC has been severely limited by its inherently low efficiency. 
Various strategies for UC enhancement have been developed within the last decade~\cite{Li2013a} involving broadband absorption~\cite{Zou2012}, triplet-triplet annihilation~\cite{Zhao2011,Simon2012}, high excitation irradiance~\cite{Zhao2013}, enrichment of molecular antenna triplets~\cite{Garfield2018}, phonon mediated enhancement~\cite{Zhou2018}. 
Among different enhancement strategies, a metal-enhanced UC~\cite{Wu2014,Dong2018a,Qin2021} has been considered as the most promising. 
The values of UC enhancement factor of $\sim 10^4$ have been reported for Au nanorods~\cite{Zhan2015} and Ag nanocubes~\cite{Wu2019d}.
Alternatively, micrometer-sized all-dielectric structures were shown to provide up to $\sim 10^5$ UC enhancement~\cite{Liang2019} with implications to UC lasers in bare~\cite{Fernandez-Bravo2018} and TiO$_2$-coated~\cite{Liu2020d} polystyrene microspheres.
Recent reviews on plasmon-enhanced upconversion~\cite{Wu2014,Dong2018a,Qin2021} have concluded that UC enhancement of $\sim 10^4$ is quite an achievement already for an emitter with rather low intrinsic quantum yield, $q_0$ (for more details on intrinsic quantum yield dependence see Ref.~\citenum{Rasskazov22Plas}).

\begin{figure}[h!]
\centering
\includegraphics[width=6in]{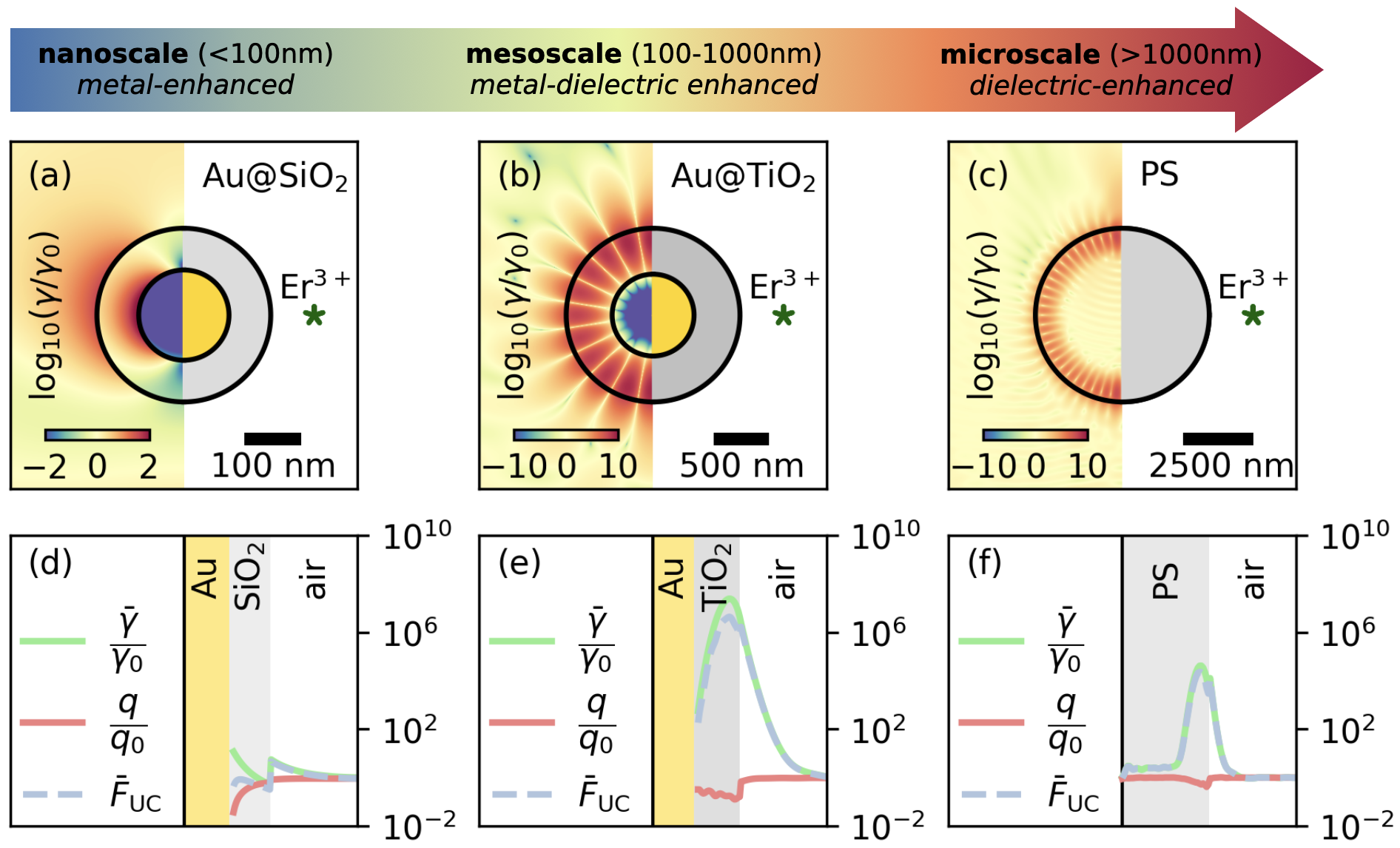}
\caption{A graphical illustration of a mesocavity which is set to occupy an intermediary region between nanoscale ($\lesssim 100$ nm) and microscale ($\gtrsim 1$ $\mu$m).
Top panels show electric field patterns at $\ld_{\rm exc}=976$ nm for the most representative cases of large UC enhancements:
(a) conventional metal-enhanced UC via Au@SiO$_2$ nanosphere (core radius $r_c=80$ nm and shell thickness $t_s=74$ nm), 
(b) metal-dielectric enhanced UC via Au@TiO$_2$ mesosphere ($r_c=374$ nm and $t_s=419.8$ nm), and 
(c) dielectric-enhanced UC via polystyrene (PS) microsphere ($r_c=3110$ nm).
Bottom panels (d)--(f) display corresponding upconversion (``UC''), excitation rate (``$\bar\gm/\gm_0$'' at $\ld_{\rm exc}=976$ nm), and quantum yield (``$q/q_0$'' at $\ld_{\rm ems}=540$ nm) enhancements of Er$^{3+}$ dipole emitter within the dielectric shell (particle) and in the vicinity of particles.
Intrinsic quantum yield, $q_0$, is set to 50\% in all simulations performed by our public MATLAB code Stratify~\cite{Rasskazov20OSAC}.
}
\label{fig:nano-meso-micro}
\end{figure}
In this work, we demonstrate a novel \textit{mesocavity} (MC) concept for the UC enhancement. 
The MC concept is based on a simple yet counter-intuitive idea of covering metal nanoparticles (NPs) with unusually thick ($\gtrsim 100$ nm) high-index dielectric shells~\cite{Rasskazov21JPCL} (Figure~\ref{fig:nano-meso-micro}). 
Our motivation comes from recent application of the mesocavity concept to \textit{fluorescence enhancement} by means of Au@high-index dielectric core-shells~\cite{Rasskazov21JPCL}. 
In the latter case, extreme fluorescence enhancement factors $F\gtrsim 3000$ for emitters located on the surface or in the interior of the shell have been observed (already for emitters with 100\% intrinsic quantum yield)~\cite{Rasskazov21JPCL}.
Here, as the proof of principle, 
we demonstrate that record high values of UC enhancement of $\gtrsim 10^6$ ($\gtrsim 10^8$) on the surface (within dielectric shell) are possible already in the case of a Au@TiO${}_2$ mesocavity.

In what follows, we elucidate
the concept of a mesocavity. 
Then we provide a summary of the theory of upconversion enhancement employed in our simulation. 
After presenting numerical simulation results of upconversion enhancement in mesocavities, a comparison with some recent concepts~\cite{Yang2017c} and upconversion enhancement results~\cite{Xomalis2021} are discussed.

\section{Mesocavity}
In this work, a mesocavity is embedded in a homogeneous medium with a refractive index $n_h$, which is set to be air ($n_h=1$).
For a particular combination of materials and geometrical parameters of metal-dielectric core-shell NPs, a ``mesocavity'' with a resonance at $\ld_{\rm exc}$ can be realized. 
As an example, we focus below on the core-shell mesocavities with a gold core (with refractive index $n_c$~\cite{Johnson1972}) surrounded by a high-index dielectric shell of $n_s=2.7$ representing the value of TiO${}_2$.
\begin{figure}
    \centering
    \includegraphics[width=5.5in]{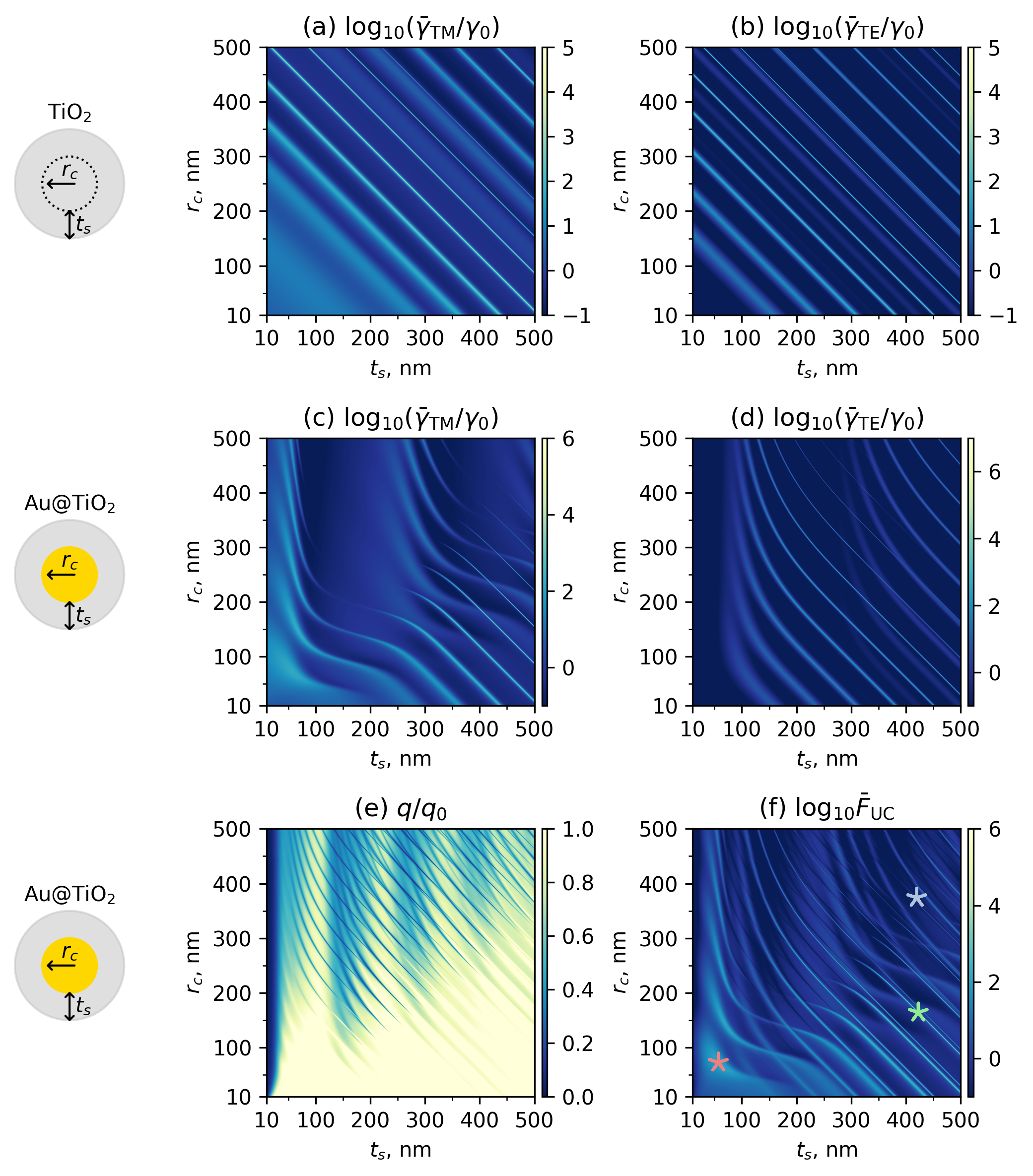}
    \caption{Averaged UC excitation rate enhancement $\langle |\tl\vE|^4 \rangle$ 
    on the outer surface of homogeneous TiO$_2$ (a,b) and core-shell Au@TiO$_2$ sphere  (c,d) for the respective TM and TE modes.
    Homogeneous TiO$_2$ sphere in (a,b) is presented as a core-shell sphere (with TiO$_2$ core and TiO$_2$ shell) for a direct comparison with Au@TiO$_2$ core-shell sphere in (c,d).
    (e) Quantum yield and (f) resultant averaged UC enhancement for the core-shell Au@TiO$_2$ sphere.
    Stars in (f) represent configurations
    $r_c=72.7$ nm, $t_s=57$ nm (red),
    $r_c=164.2$ nm, $t_s=422.6$ nm (green),
    $r_c=374$ nm, $t_s=419.8$ nm (blue) highlighted in detail in Figure~\rf{fig:UC_inter}.
    }
    \label{fig:flds}
\end{figure}
The lessons to be learned from Figure \ref{fig:flds} are that the presence of a metal core
(i) does not deteriorate the high values of quality factor ($Q$-factor) of all dielectric resonances and 
(ii) bends the contours of all dielectric resonances in the $(t_s,r_c)$ parameter space towards smaller valuers of $r_c$ (TM resonances significantly more than TE resonances).
Therefore, the use of a metal core in mesocavity enables one to access a kind of all dielectric high-$Q$ resonances with overall smaller cavity radius ($r_s=r_c+t_s$).
This explains why the radius ($r_c=3110$ nm) of a polysterene sphere shown in Figure~\ref{fig:nano-meso-micro}(c) can be significantly reduced down (to $r_s\approx 794$ nm) while maintaining high UC values.
In spite of a thick shell, a metallic core is indispensable, because nothing comparable happens for analogous resonances in a comparable homogeneous dielectric spheres (see Figure~\ref{fig:flds}), which is consistent with Refs~\cite{DeDood2001a,Sun2014b}. 
Compared to the case of a homogeneous dielectric sphere, the metal core plays the role of a reflecting surface pushing the electric field intensity from the core interior into the outer shell part~\cite{Rasskazov21JPCL}, thus creating kind of pseudo-cavity~\cite{Patra2021}.

Our mesocavity concept can be seen as facilitating a beneficial synergy of plasmonic~\cite{Tovmachenko2006,Stockman2011,Osorio-Roman2014} and  whispering gallery-mode resonances (WGR)~\cite{Braginsky1989} functionalities within an intermediary region between nanoscale ($\lesssim 100$ nm) and microscale ($\gtrsim 1$ $\mu$m) (Figure \ref{fig:nano-meso-micro}). 
Plasmonics of individual truly nanoscopic particles is essentially captured by their dipole properties, with quadrupolar resonance hardly playing any significant role. 
The mesocavity concept provides  a versatile tool of accessing more interesting physics by being able to make use of a number of first few multipoles ($\ell\lesssim 10$).
A dipole emitter located on top or within an outer region of a shell of such mesocavity is well separated from the metal surface to substantially mitigate its nonradiave losses (the normalized nonradiative decay rate $\tilde{\Gm}_{\rm nrad} \ll 1$)~\cite{Rasskazov21JPCL}. 
At the same time, large electromagnetic field enhancement is engineered sufficiently far away from a metal surface, which is out of reach for conventional ``shell-isolated'' {\em metal-enhanced} processes~\cite{Tovmachenko2006,Osorio-Roman2014}.

\section{UC enhancements in a mesocavity}
Present study of upconversion enhancement is limited to the most common case of electric dipole (ED) transitions. 
UC molecule, being located in a vicinity of a particle, experiences a modification of both excitation and emission rates.
An enhancement factor of upconversion bears a lot of similarities with enhancement factor of the ED fluorescence~\cite{Bharadwaj2007}. The sole difference in that the UC excitation rate, $\gamma\propto \lvert \vE\rvert^4$, is proportional to the {\em square} of the intensity of the electric field compared to its {\em linear} dependence on the intensity for fluorescence~\cite{Bharadwaj2007}. One has formally~\cite{Auzel2004,Wu2014,Qin2021}:

\begin{equation}
    F_{\rm UC} = \dfrac{\gm}{\gm_0} \times \dfrac{q}{q_0} \ ,
 \label{eq:fuc}
\end{equation}
where $\gm$ is the UC excitation rate, $q$ is the quantum yield, and the subscript ``0" indicates the respective quantity in the free space.
As alluded to earlier, the excitation rate of UC, being a two-photon process, is according to the Fermi's golden rule proportional to the {\em square} of the intensity of the electric field in the location of the UC molecule, $\gamma\propto \lvert \vE\rvert^4$. In the presence of a particle, quantum yield of UC emitter is modified due to radiative and non-radiative decay rates~\cite{Bharadwaj2007}:
\begin{equation}
    q  = \dfrac{\tilde{\Gm}_{\rm rad}}{\tilde{\Gm}_{\rm rad} + \tilde{\Gm}_{\rm nrad} + (1-q_0)/q_0 } \ ,
\end{equation}
where the quantities with tilde are normalized to the {\em radiative} decay rate in free space (e.g. $\tilde{\Gm}_{\rm (n)rad}=\Gm_{\rm (n)rad}/\Gm_{{\rm rad};0}$).
Explicit expressions for radiative and non-radiative decay rates for a case of a multilayered sphere have been presented in Ref.~\citenum{Moroz2005}.

Simulations are performed by freely available MATLAB code Stratify~\cite{Rasskazov20OSAC}.
For ED located at the radial distance $r$ from the mesocavity origin, the UC enhancements are averaged in a sense that: 
\begin{itemize}
\item decay rates $\Gm_{\rm (n)rad}$ at any particular position {\bf r} are averaged over all possible ED orientations.
In other words, $\Gm_{\rm (n)rad} = (\Gm^\perp_{\rm (n)rad} + 2\Gm^\parallel_{\rm (n)rad})/3$, where superscripts ``$\perp$'' and ``$\parallel$'' denote radial and tangential orientation of ED, respectively;
\item the excitation rate, $\gm$, is averaged over the spherical surface of radius $r$ according to closed-form analytical solution~\cite{Rasskazov19JOSAA}.
\end{itemize}
The value of intrinsic quantum yield is set to $q_0=0.5$ (maximum value for a two-photon process~\cite{Wu2014}) and held constant throughout this work.

\begin{figure}
    \centering
    \includegraphics{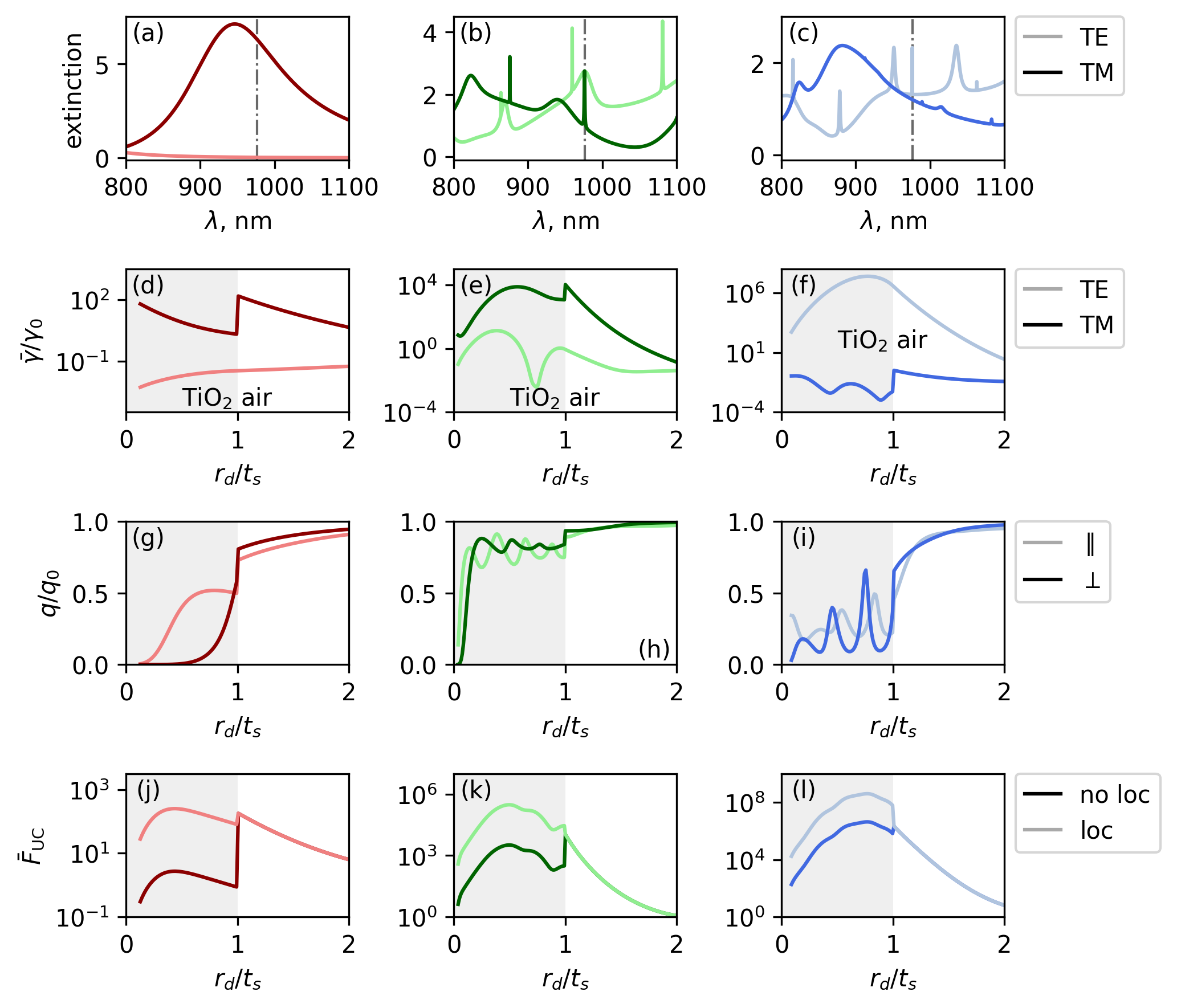}
    \caption{
    Results for the configurations of core-shell Au@TiO$_2$ spheres with
    $r_c=72.7$ nm, $t_s=57$ nm (left column, $Q\approx4.3$ for TM mode of $l=1$ order),
    $r_c=164.2$ nm, $t_s=422.6$ nm (middle column, $Q\approx1100$ for TM mode of $l=6$ order),
    $r_c=374$ nm, $t_s=419.8$ nm (right column, $Q\approx9745$ for TE mode of $l=10$ order).
    First row: extinction spectra decomposed in the respective TE and TM polarizations,
    showing the effect of mesocavity resonances in the panels (b) and (c). 
    Second row: excitation factors for TE and TM polarizations (without local field correction).
    Third row: quantum yield as function of distance for emitter with tangential ($\parallel$) and radial ($\perp$) orientation.
    Fourth row: resultant UC enhancement with and without taking into account the Lorentz local-field correction of eq~\rf{Lvc}.}
    \label{fig:UC_inter}
\end{figure}

Figure~\ref{fig:flds}(f) shows that \textit{averaged} UC enhancement factors can reach up to $\approx10^6$ values for emitters located on top of MCs, due to simultaneously large values of electric field enhancement (Figure~\ref{fig:flds}(c)--(d)) and quantum yield (Figure~\ref{fig:flds}(e)).
Although the presence of lossy components appears counterintuitive, resonance $Q$-factors were demonstrated~\cite{Rasskazov21JPCL} with the values of $\sim 10^4$ between that of typical plasmonic structures~\cite{Stockman2011} ($Q\lesssim 10^2$) and WGMs in large homogeneous silica spheres~\cite{Braginsky1989} ($Q\ge 10^8$). 
It is shown in Figure~\ref{fig:UC_inter}, where three configurations of Au@TiO$_2$ spheres are considered in details.
$Q$-factor of typical plasmonic NP is relatively low $\approx4.3$ for a dipolar mode ($l=1$, Figure~\ref{fig:UC_inter}(a)), while for MCs it reaches values of up to $\approx1100$ and $\approx9745$ for $l=6$ and $l=10$ multipoles (Figures~\ref{fig:UC_inter}(b) and (c), respectively).
Mesocavity $Q$-factors are thus comparable to the highest $Q$-factors of all-dielectric structures of similar size~\cite{Huang2021} and exceeding those of ultra-high-$Q$ resonances in plasmonic metasurfaces~\ct{Bin-Alam2021}. 

When the polarizability of a guest dipole is sufficiently low compared to that of
the host, the spontaneous emission lifetime obeys the {\em real-cavity} (also known as empty-cavity) model. This also applies to the case when an emitter expels dielectric medium and creates a real (tiny) cavity therein. Both cases are covered by the so-called substitutional case of Ref.~\citenum{Vries1998}.
For pure systems constituted of only one kind
of atom, or when the polarizability of the guest is the same as that of the host, the interstitial case of Ref.~\citenum{Vries1998} described by the virtual-cavity model applies. This is also the case of rare earth emitters implanted in dielectrics by ion beam deposition~\cite{DeDood2001a} (but not the case of rare earth emitters embedded in various organic complexes~\cite{Rikken1995,Duan2005}).
A convincing compilation of experimental results confirming the theoretical reasoning by Vries et al.~\cite{Vries1998} has been provided in Refs.~\citenum{Duan2005,Aubret2019}.

In the interstitial case described by the virtual-cavity model,
the local field, $\vE_{loc}$, felt by emitters in the presence of a macroscopic field $\vE$ is $\vE_{loc}=L_{vc} \vE$, where $L_{vc}$ is the so-called Lorentz local-field correction (see Supplementary Material for a general derivation of the Lorentz local-field correction from the Maxwell's equations):
\bg
1< L_{vc}= \fr{\veps+2}{3}\cdot
\lb{Lvc}
\eg
The Lorentz local-field correction in eq~\ref{Lvc} is particularly interesting for high-index dielectric hosts, because $L_{vc}$ linearly increases with the host dielectric constant and it can become large: $L_{vc}=3.1$ for the shell refractive index of $n=2.7$. 
Given that the excitation process is $\propto|\vE_{loc}|^4$, the Lorentz local-field correction can provide an additional enhancement factor of $\approx 92$ for $n=2.7$ for the emitters embedded within high-index shell. 
The latter is illustrated in Figure \ref{fig:UC_inter}.
Significantly, rare earth emitters implanted in dielectrics by ion beam deposition
fall under the interstitial case~\cite{DeDood2001a}, which provides a promising and experimentally feasible way of achieving high UC rates.

\section{Discussion}
It is expedient to compare our approach with that of Yang et al \cite{Yang2017c}, who theoretically demonstrated that metal losses can be successfully mitigated with high refractive index dielectric particles on metallic films. 
Our mesocavity concept shares three main features with the approach of Yang et al. \cite{Yang2017c} in  achieving resonances with high $Q$, which are, counterintuitively, (i) the necessity of a lossy metal component (a metallic core in our case vs planar metal film of Yang et al \cite{Yang2017c}), (ii) a dielectric gap (a dielectric shell in our case), and (iii) high refractive index of a dielectric. The crucial differences of our approach to that of Yang et al.~\cite{Yang2017c} is:  
(i) we make use of free individual particles, hence not requiring any substrate, like the planar metal film covered by a dielectric spacer layer of Yang et al.~\cite{Yang2017c} and, probably because of using particles, 
(ii) the dielectric shell should be sufficiently thick (well above $\sim 30$ nm, a typical upper limit on emitter-metal separation in metal enhanced fluorescence, and much larger than the gap of 2-5nm of a dielectric cylinder on a semi-infinite metal surface of Yang et al. \cite{Yang2017c}). 

Recently, Xomalis et al.~\cite{Xomalis2021} converted approximately 10-$\mu$m-wavelength mid-infrared (MIR) incoming light to visible light by surface-enhanced Raman scattering (SERS) in doubly resonant antennas that enhanced upconversion by more than $\sim 10^{10}$. They tailored their structure to have one resonance at the predetermined pump frequency and the other at the MIR frequency, and the Stokes shift was fine-tuned to the MIR photon they intended to detect. Very much like in the work of Yang et al.~\cite{Yang2017c}, Xomalis et al.~\cite{Xomalis2021} design required the use of a planar nanoantenna array and not of individual particles as in the present work.
While their ingenious concept~\cite{Xomalis2021} is well suited for the MIR light upconversion, it seems difficult to apply the concept for the present case of near-infrared (NIR) upconversion. Contrary to that, the proposed mesocavity concept proposed is well suited for the NIR UC enhancement of the order of  $\gtrsim 10^8$, which could be advantageous for applications in biology~\cite{Chen2014,Chen2015,Zhou2015b,Wu2016,Wang2018d}, imaging~\cite{Huot2016}, solar energy harvesting~\cite{Wang2014,Jang2016,Meng2016,Li2018}, and nanoscopy~\cite{Liu2017f,Zhan2017}.

\section{Conclusions}
We have demonstrated the mesocavity concept based on a simple yet counter-intuitive idea of covering metal nanoparticles with unusually thick ($\gtrsim 100$ nm) high-index dielectric shells~\cite{Rasskazov21JPCL} for the UC enhancement.
A mesocavity facilitates a beneficial synergy of plasmonic and WGR functionalities within an intermediary region between nanoscale ($\lesssim 100$ nm) and microscale ($\gtrsim 1$ $\mu$m) (Figure \ref{fig:nano-meso-micro}).
On the example of a core-shell mesocavity with a gold core surrounded by a high-index dielectric TiO${}_2$ shell with $n_s=2.7$, the UC enhancement on mesocavity surface can be as large as $\gtrsim 10^6$. 
Including the Lorentz local-field correction, an additional enhancement factor for UC of $\approx 92$ for emitters embedded in the high-index dielectric shell with refractive index $n=2.7$ is possible, resulting in the total upconversion enhancement of $\gtrsim 10^8$, which is three orders of magnitude larger than it has ever been reported. At least the same results are expected for 
core-shell mesocavities with cores made of other plasmonic materials (Al, Ag, Au, Mg)~\cite{Doiron2019}, and with TiO${}_2$ being replaced with comparable high-index dielectric materials (TiO${}_2$, Ta${}_2$O${}_5$, ZnS, Nb${}_2$O${}_5$, SiC, GaP, amorphous Si)~\cite{Baranov2017b}, because the achieved UC enhancement correlates with the shell refractive index.
Numerous commercially available dyes implies an impressive variety of possible NP-dye combinations. 

\begin{suppinfo}
A general derivation of the Lorentz local-field correction directly from the Maxwell's equations.

\end{suppinfo}

\bibliography{references}

\end{document}


\usetagform{supplementary}

\section{Local field corrections}
The local field within the cavity, $\vE_{loc}$, differs from an applied macroscopic field, $\vE$ by a local-field correction factor $L$, $\vE_{loc}=L \vE$.
Local field corrections inside dielectrics exhibit the well-known (i) {\em real}, or {\em empty-cavity}, and (ii) {\em virtual-cavity}, or {\em Lorentz local-field}, factors for {\em substitutional} and {\em interstitial} atoms, respectively~\cite{Vries1998}. 

The real cavity model assumes that (i) the atom is at the 
center of the cavity and (ii) the cavity itself has no other material, i.e. it is empty. The resulting ratio of local and macroscopic fields is
given by [cf. electrostatic result \rfs{elocinc} below]
\bg
1 <  L_{rc}= \fr{3\veps}{2\veps+1} < \fr32 \cdot
\lb{Lrc}
\eg
The substitutional case occurs prevalently for impurity atoms \cite{Vries1998} and rare-earth emitters embedded within different organic complexes~\cite{Duan2005}, whenever the emitter embedded in a dielectric host expels the dielectric media and creates there a real tiny cavity.
As pointed out by B\"ottcher based on the initial work of Onsager in 1933,
the local-field correction factor depends also on the polarizability $\chi$ of the molecule placed in the cavity. The real cavity model applies
when the polarizability of the guest dipole is sufficiently low compared to that of the host so that the reaction field caused by the induced dipole acting on the cavity can be neglected~\cite{Aubret2019}. 

The {\em virtual} cavity model assumes a uniform distribution of material within and outside the cavity~\cite{Aspnes1982,Vries1998,Dolgaleva2012,Aubret2019}. 
The corresponding local-field correction is then known as the Lorentz local-field correction, eq~3, 
\bg
1< L_{vc}= \fr{\veps+2}{3}\cdot
\lb{Lvcs}
\eg

For pure systems constituted of only one kind
of atom, or when the polarizability of the guest is the same as that of the host, the interstitial case of Ref.~\cite{Vries1998} described by the virtual-cavity model applies. This is also the case of rare earth emitters implanted in dielectrics by ion beam deposition~\cite{DeDood2001a}.

One can notice a significant difference between the respective local-field corrections \rfs{Lrc} and eq~3: whereas $L_{rc}$ is strictly bounded by $3/2$ from above, the Lorentz local-field correction \rfs{Lvcs} increases indefinitely with increasing $\veps$ and is, in principle, unbounded from above. In general, $L_{rc}$ represents a lower bound for the local field factor~\cite{Aubret2019}.

The textbook derivation of local-field corrections is usually performed within quasi-static approximation~\cite{Aspnes1982,Vries1998,Dolgaleva2012,Aubret2019}. Let us consider a sphere with dielectric constant $\veps_s$ embedded in a host characterized by dielectric constant $\veps_h$. Irrespective of the units used, one finds for the sphere in the electrostatic case
\bg
\vE_{loc} = \fr{3\veps_h}{\veps_s + 2\veps_h}\, \vE_{inc}.
\lb{elocinc}
\eg
The above familiar electrostatic result illustrates the necessity of applying local field corrections when estimating the local excitation field $\vE_{loc}$ felt inside a {\em real} cavity by atoms and molecules in the presence of a macroscopic field $\vE_{inc}$.

Less known is a general derivation of the Lorentz local-field correction directly from the Maxwell's equations.
In general, i.e. beyond the electrostatic approximation, 
the origin of the local-field factors is that the conventional electric dyadic Green's function is {\em not} sufficient to determine the correct value of $\vE({\bf r})$ at source points~\cite{Yaghjian1980}. One has
\bg
\vE({\bf r}) =i\om\mu_0\, \lim_{\dt\rar 0} \int_{V_j-V_\dt} \vG({\bf r},{\bf r}')\cdot\vJ ({\bf r}')\, dV'
+ \fr{\vL\cdot\vJ({\bf r})}{i\om\veps_0 },
\lb{elfvl}
\eg
where $V_\dt$ is an excluded volume, i.e. a cavity comprising the observation point {\bf r}, $\vG$ is the Green's function to the equation
\bg
[\vnab\times\vnab\times - k^2] \vG = \dt({\bf r}-{\bf r}'){\bf I},
\lb{vgdef}
\eg
and $\vL$ is an extra dyadics~\cite{Yaghjian1980}
\bg
\vL=\fr{1}{4\pi}\oint_{S_\dt} \fr{\vn\otimes \ve_{R'}}{R'^2}\, dS'.
\lb{vlform}
\eg
Here $\vn$ is the unit normal pointing out of the principal volume 
and $\ve_{R'}$ is the unit vector pointing from ${\bf r}$ to ${\bf r}'$.
The integral on the rhs of \rfs{elfvl} is an {\em improper} nonconvergent integral in the sense of Kellogg~\cite{Kellogg1967}, i.e. it is necessary to restrict the shape of $V_\dt$ in order to obtain a limit when $V_\dt\to 0$. 
Nevertheless, although each of the two contributions on the rhs of \rfs{elfvl} does individually depend on the shape of of excluded volume $V_\dt$, the rhs of \rfs{elfvl} as a whole is independent of the shape of $V_\dt$.
For arbitrary principal volumes and time harmonic fields,
$\vL$ can be concisely interpreted physically as a generalized depolarizing dyadics yielding the ``local field" $\vE_0 + \vL\cdot\vJ/\veps_0$ of electrostatics. Conversely, the  depolarizing factors for an ellipsoid and the generalization to arbitrary shaped holes in or bodies of uniform volume sources can be found by the formula \rfs{vlform}.

Suppose one were to measure the electric field at a point within 
an enforced current distribution by removing an infinitesimally
small volume $V_\dt$ of current and inserting an ideal point probe therein.
Then the $\vL$ term determines the perturbation in electric field
caused by the hypothetical measurement scheme of removing 
an infinitesimally small volume $V_\dt$ of enforced uniform current. 
Indeed, the measured field would then be given by \rfs{elfvl} but without the 
$\vL$ term, since the enforced current at this point has been removed.
The measured local field would then be~\cite{Yaghjian1980}
\bg
\vE({\bf r}) - \fr{\vL\cdot\vJ}{i\om\veps_0 },
\lb{elfvloc}
\eg
and would depend upon the shape of the infinitesimal volume and its relative 
position and orientation with respect to the point probe.
Provided that $\vJ$ here is merely an enforced polarization current, and assuming harmonic $e^{-i\om t}$ time dependence,
\bg
\vJ=\pa_t \vP=-i\om \vP,
\eg
one obtains in the case of dielectrics
\bg
\vJ({\bf r}) =- i\om\veps_0  (\veps({\bf r}) - 1) \vE({\bf r}),
\lb{vJE}
\eg
where $\veps$ is the relative dielectric contrast, $\veps({\bf r}) =\veps({\bf r})/\veps_h$. The extra dyadics is known
for a variety of different principal volumes. For instance, $\vL={\bf I}/3$ for a {\em sphere}
and a {\em cube} (see Table 1 in Ref.~\cite{Yaghjian1980}). 
According to \rfs{elfvloc}-\rfs{vJE}, the measured local field is then
\bg
\vE_{loc}({\bf r})=\vE({\bf r}) + \fr{\vP}{3\veps_0 }=\vE({\bf r})
\left(1 + \fr{\veps({\bf r}) - 1}{3}\right)
=\fr{\veps({\bf r})+2 }{3} \, \vE({\bf r}).
\lb{elfvlocm}
\eg
This way the quasi-static Lorentz factor, or the Lorentz local-field correction in eq~3, is recovered.

\bibliography{references}

%% file: new.tex
\newcommand{\be}{\begin{equation}}
\newcommand{\ee}{\end{equation}}
\newcommand{\bg}{\begin{equation}}
\newcommand{\eg}{\end{equation}}
\newcommand{\bdm}{\begin{displaymath}}
\newcommand{\edm}{\end{displaymath}}
\newcommand{\bea}{\begin{eqnarray}}
\newcommand{\eea}{\end{eqnarray}}
\newcommand{\beas}{\begin{eqnarray*}}
\newcommand{\eeas}{\end{eqnarray*}}
\newcommand{\ba}{\begin{array}}
\newcommand{\ea}{\end{array}}

\newcommand{\bfg}{\begin{figure}}
\newcommand{\efg}{\end{figure}}

\newcommand{\tl}{\tilde}

\newcommand{\fr}{\frac}

\newtheorem{lm}{Lemma}
\newtheorem{cl}{Corollary}
\newtheorem{df}{Definition}

\newcommand{\blm}{\begin{lm}}
\newcommand{\elm}{\end{lm}}
\newcommand{\bcl}{\begin{cl}}
\newcommand{\ecl}{\end{cl}}
\newcommand{\bdf}{\begin{df}}
\newcommand{\edf}{\end{df}}
\newcommand{\brk}{\begin{rm}}
\newcommand{\erk}{\end{rm}}

\newcommand{\lb}{\label}

\newcommand{\veps}{\varepsilon}

\newcommand{\ld}{\lambda}

\newcommand{\gm}{\gamma}

\newcommand{\Gm}{\Gamma}

\newcommand{\ct}{\cite}
\newcommand{\rf}{\ref}

\newcommand{\vE}{{\bf E}}

